\input epsf.tex

\magnification\magstep 1
\font\ff = cmr9            
\font\sy = cmsy10          
\font\ref = cmr12          
\vsize=1.05\vsize
\voffset=-15true mm
\tolerance=10000

\newcount\footnum
\footnum=0
\def\fta{\relax}
\let\footold=\footnote
\def\footnote#1#2{\global\advance\footnum by 1
\footold{#1${}^{\the\footnum}$}{#2}}

\centerline {\bf RESONANCE TYPE INSTABILITIES}
\centerline {\bf IN THE GASEOUS DISKS}
\centerline {\bf OF THE FLAT GALAXIES}
\centerline {\bf II. The stability of solitary vortex sheet}
\medskip
\centerline {C.M.Bezborodov and V.V.Mustsevoy}
\bigskip

{\ff
Linear stability analysis of the axisymmetric interface of velocity
and density discontinuity in rotating gaseous disk has been performed
numerically and analytically. Physical mechanisms leading to development
of centrifugal and Kelvin--Helmholtz instability at the kink has been
analysed in detail. In the incompressible limit it has been shown
in the first time such areas in the parameter space that
Kelvin--Helmholtz instability is stabilized by the density kink.
This effect is caused by both specifical angular momentum conservation
and buoyancy. The possibility of application of obtained results to the
stability analysis of the gaseous disks of the real flat galaxies is
discussed.
}
\medskip  
{ \it Introduction.}
\medskip  

The interest to a problem of stability of gaseous galactic disks
is caused by the hydrodynamic concept of an origin of spiral
structure of galaxies proposed in 1972 by A.M.~Fridman (Fridman 1978).
According to this concept the galactic spiral arms represent waves
of density, arising and then reaching nonlinear stage as result
of hydrodynamic (non-Jeans) instabilities development in a gaseous disk.

The most probable applicants for a role of the spiral structure
generator are thus Kelvin--Helmholtz and centrifugal instability.
The development of these instabilities on a shift layer in rotating
cylindrical and disk configurations of gas was already investigated
many times (see Morozov et al. 1975, 1976a, 1976b, Morozov
1977, Landau 1944). In according to Morozov (1979) and Torgashin
(1986) development of hydrodynamic instability in model with smooth
VJ-like rotation curve (i.e. having significant velocity fall
between solid-body rotating and plato parts) was considered, and the
jump of angular speed was concurrent to jump of density.

However, because of impossibility of analytical study the solution was
found numerically, and from multiparametricity of a problem it was not
possible to trace dependence of instability properties (growth rate)
from parameters of model in details. Nevertheless we note that
Morozov (1989) reported on instability weaking with growth of
density contrast between inner and outer disk areas and about
minority of influence of sound speed contrast. In all the other
quoted works the analytical asymptotic solutions were received
in uniform density models\footnote {\fta} {\ff It is impossible
to bypass work of Fridman (1989), where the deep analysis of
gradient instabilities in a gaseous disk in the application
to the hydrodynamic concept of spiral structure origin is spent.
Unfortunately in the paragraph devoted to stability of velocity
jump concurrent with density jump, faulty boundary condition
(then corrected by Friman \& Khoruzhii (1993)) was used,
but investigation of growth rates of instability was not carried out.}.

We continue study of sharp speed jump stability, concurrent on radius
with density jump, and promote it in discontinuous non-homentopic model.

In the present part of the article we shall use the references to
the formulae from a Part I, adding in this case  symbol ``1.''
before number of formula.

\medskip 
{ \it 1. Model and dispersion equation.}
\medskip 

The most simple and convenient for the analysis model with density
profile discontinuity cannot be polytropic (Morozov \& Mustsevoy 1989).
It is connected that at $p(r) = K\sigma^\gamma (r) $, where $K = const$,
the density jump existence is equivalent to jump of pressure.
The same takes place for three-dimensional quantities too.
By virtue of a condition of radial forces balance:
$$
   {V_{\varphi}^2 \over r} =
      {1 \over \rho} {dP \over dr} + {d \Psi \over dr}, \eqno (1)
$$
where $V_\varphi$ is gas rotation linear speed, $\Psi$ is gravitational
potential, the pressure jump with necessity entails jump of gravitational
potential (the potential of double layer).

The data on distributions of galactic clouds velocities dispersion are
practically absent; we shall use non-homentopic model with density
jump but uniform pressure, that imposes relation on three-dimensional
parameters of gas at both sides from jump:
$\rho_{ex} / \rho_{in} = {c_s^2}_{in} / {c_s^2}_{ex} $
(the index ``{\it in}'' designates quantity from the inner
side of jump and index ``{\it ex}'' with outer one).

Considering told we research dynamics of distur\-bances of small
ampli\-tude on a background of thin three-dimensional non-selfgravitating
non-homentopic disk unlimited on radius being in external stationary
axisymmetic potential and consisting of ideal non-viscous gas.
Following a standard linearization procedure we present all
describing disk magnitudes as follows:
$$
	f(r,\varphi, z, t) = f_0 (r, z) + \tilde f (r,\varphi, z, t),
	\ {\rm where} \ |\tilde f|\ll f_0. 			\eqno (2)
$$
The equilibrum axisymmetic stationary condition described by the radial
balance equation (1) is charac\-te\-rized by the following parameters
distri\-bu\-tions:
$$\eqalign {
\rho_0&(r, z) = \rho_{in}(z) +
	(\rho_{ex}(z) - \rho_{in}(z)) \,\theta (r-R), \cr
c_s&(r, z) = {c_s}_{in}(z) \left[ 1 +
      \left( \sqrt {\rho_{in}(z) \over\rho_{ex}(z)} - 1 \right)
      \theta (r-R) \right], \cr
\Omega&(r, z) = \Omega_{in}(z) +
	(\Omega_{ex}(z) - \Omega_{in}(z)) \,\theta(r-R), \cr
P_0&(r, z) = c_s^2 (r, z) \rho_0(r, z) /\gamma = const (z), \cr} \eqno (3)
$$
where $\theta(x) $ is centered Heaviside function,
$\theta(x) = 0$ at $x < 0$, $\theta (x) = 1$ at $x > 0$ and
$\theta(0) = 1/2$, and $\gamma$ is adiabatic index.

We do not use the Poisson's equation determining gravitational
potential because dealing with model problem we shall hereafter
implicitly set necessary for us form of $\Psi_0(r, z)$ believing
condition (1) for given distributions (3) and condition
of hydrostatic balance along $z$--coordinate holding.

Because of uniformity of model parameters on azimuthal angle
$\varphi$ and time $t$ the solution of gas dynamics linearized
equations system is searched by decomposition on normal modes.
Thus separate Fourier-harmonics of a kind
$\tilde f \propto f (r, z) \exp (i m \varphi - i \omega t)$
evolution is further investigated. As Morozov (1989) show,
for non-homentopic in a disk plane model
$(P_0(r, z) / {\rho_0^\gamma(r, z)} \ne F(z))$ it is described
by the system of equations:
$$
  {dp\over dr} = \left({2m\Omega\over r\hat\omega}
  + {P_0^\prime\over\rho_0  c^2_s} \right) p
  + \rho_0  \left[\hat\omega^2 - \kappa^2 +
  {P_0^\prime\over\rho_0 } \left({P_0^\prime\over c_s^2\rho_0 }
  - {\rho_0^\prime\over\rho_0 } \right) \right] \xi, 		\eqno (4)
$$
$$
  {d\xi\over dr} = \left({m^2 c_s^2 \over r^2 \hat\omega^2} - 1 \right)
  {P\over c_s^2 \rho_0 } - \left({2m\Omega\over r\hat\omega} +
  {1\over r} + {P_0^\prime\over c_s^2 \rho_0 } \right) \xi. 	\eqno (5)
$$
Here designations of a Part I are kept, i.e. $p$ is disturbance
of pressure, $\xi$ is perturbed radial Lagrange displacement,
correlated to disturbance of a radial speed component by relation:
$ {d\tilde\xi / dt} = -i (\omega - m\Omega) \tilde\xi = -i\hat\omega\tilde\xi
= \tilde v_r$, $\hat\omega$ is Doppler frequency,
$\kappa^2 = 2\Omega (2\Omega + rd\Omega/dr) $ is squared epicyclic
frequency, $c_s$ is sound speed, the prime stands for differentiation
on radius.

The system (4)--(5) adequately describes dynamics of small
distur\-bances in such objects as:

$\bullet$ a differentially rotating cylindrical configuration of
non-dissipative non-selfgravitating ideal gas with the equation of state
$P_0 = \rho_0 c_s^2 / \gamma$ ($P_0$ and $\rho_0$ are homogeneous
along cylinder forming equilibrum pressure and density respectively,
$c_s$ is adiabatic sound speed, $\gamma$ is adiabatic index);
the description is adequate for disturbances with $v_z \equiv 0$;

$\bullet$ a thin non-dissipative non-selfgravitating disk of polytropic
gas, being in stationary external (star) gravitational potential
of a kind $\Psi(r,z) = \Phi(r) + \chi(z)$ (see Appendix I in the work
of Gorkavy \& Fridman (1994)); pinch type disturbances on
$z$--coordinate are in this case considered only, $v_r$ and $v_\varphi$
with necessity designates their average on $z$ values, and $P_0$ and
$\rho_0$ are surface pressure and density accordingly; the two-dimensional
adiabatic index $\gamma_s$ is associated to a volumetric by relation:
$\gamma_s = 3 - 4/(\gamma + 1)$ (Churilov \& Shukhman 1981).

Note that the realization of the last case is necessary for construction
of step--homogeneous two--dimension model of thin disk because for
potential of a more general form $\Psi = \Psi (r, z) $ either rotation
velocity averaging $V_\varphi = V_\varphi (r, z) $ on $z$--coordinate
will not lead to discontinuous distribution of $V_\varphi (r) $ or there
will be the radial inhomogeneities of pressure, which are necessary
for fulfilment of equilibrum forces balance. A really specified form
of potential means that the disk is quite thin for applicability
of potential decomposition in a series on small parameter $z/r$
with neglecting of all terms more senior than second-order.

As in each of areas divided by jump all equilibrum parameters are
homogeneous, the system (4)--(5) becomes identical to system
(1.1)--(1.2) and is reduced to modified Bessel equation on $p$,
the general solution of which is a linear combination of modified
Bessel functions $I_m(k_i r)$ and $K_m(k_i r)$ where a designation
is introduced
$$
k_{in; ex}^2 = {4 \Omega_{in; ex}^2 - (\omega - m\Omega_{in; ex})^2
      \over {c_s^2}_{in; ex}}. 					\eqno (6)
$$
Note that completely analogous equation is also valid for quantities
$\rho(r)$, $\eta(r) = \rho(r) /\rho_0$, {\sy P} $(r) = c_s^2 \eta(r)$.
Assuming hereafter for uniqueness of the solution $Re\, k_{in; ex} > 0$,
from a natural condition of its limitation in a disk centre and
on infinity follows:
$$ p (r) = \left\{\matrix {
A I_m (k_{in} r), & r < R, \cr
B K_m (k_{ex} r), & r> R. \cr
} \right. 							\eqno (7)
$$

Boundary conditions for the solutions on jump we determine by
integrating the equation (4), (5) on radius from $R-\varepsilon$ up to $R+\varepsilon$
with account of peculiarity of chosen model ---
$\partial P_0 / \partial r = 0$ and then tending $\varepsilon\to 0$
(see Fridman \& Horuzhii 1993 for details):
$$
      \xi (R + 0) - \xi (R-0) = 0, \eqno (8)
$$
$$
p(R+0) - p(R-0) = {\rho_{in} + \rho_{ex}\over 2} \,\xi(R)\,R(\Omega^2_{in} -\Omega^2_{ex} ).\eqno(9)
$$

Coupling of the solutions at $r = R$ results system of the equations
on factors $A$ and $B$. The condition of simultaneity of this system
represents the dispersion equation allowing to determine frequency
eigenvalues:
$$
\left|
\matrix {
\beta_{in} 	& {Q\over\mu^2} k_{in}^2 R^2 \alpha_{ex} \cr
1 	& k_{ex}^2 R^2 - {M^2\over 2\mu^2} (1 + Q) (1-q^2) \alpha_{ex} \cr
} \right| = 0. 							\eqno (10)
$$

Here the following designations are made: $M = {R\,\Omega_{in}/{c_s}_{in}}$
is Mach number, $q = {\Omega_{ex}/\Omega_{in}}$, $Q = {\rho_{in}/\rho_{ex}}$,
$\mu = {{c_s}_{ex}/{c_s}_{in}}$, $\alpha_i$ and $\beta_i$ are defined by (1.8)
and (1.9) in which the index $i$ still takes values ``$in$''
and ``$ex$'', and the top index is omitted because of coincidence
of radii of speed and density jumps. Besides let us introduce
dimensionless frequency $x=\omega/\Omega_{in}$. The existence of positive
$Im\,\omega$ (growth rate) means instability.

Note the original dispersion equation was satisfied identically for
neutral values ($Im\,\omega=0$): $\omega = (m\pm 2) \Omega_{in}$,
$\omega = (m\pm 2) \Omega_{ex}$, that is for gyroscopic modes of
fluctuations of a homogeneous solid-body rotating disk.
It is natural result, as the disturbances with such
frequencies propagate exclusively in an azimuthal direction
($k_{in}=0$, or $k_{ex}=0$) and are not ``sensitive'' to radial
inhomogeneities. Accordingly the given modes do not influence
system stability and deriving (10) we supposed that the frequency
does not take such values.

\medskip  
{ \it 2. The stability analysis.}
\medskip  

Let us obtain the approximated solutions of the dispersion equation
(10) for limiting cases of large and small compressibility of medium.
\medskip
{ \it a) The case of large compressibility} ($M \gg 1$)

By applying representation of arguments of Bessel functions through
dimensionless parameters note the module of argument is dictated
first of all by Mach number $M$. Assuming therefore
$ |k_{in}R|\gg m,\ |k_{ex}R| \gg m$, from (10) we find:
$$
x \simeq {1\over\mu + Q} \left\{ m(Q + q\mu)+{i\over 2}M(1+Q)(1-q^2)
\right\}.							\eqno (11)
$$
See $|x|\sim M$, $|k_{ex}R|\sim M^2$, $|k_{in}R|\sim M^2$,
therefore the assumptions made earlier are justified backdating.

For a case of constant density from (11) we get result obtained
by Morozov (1977):
$$
\omega \simeq {\Omega_{in}\over 2} \left[m (1 + q)
	+ iM (1-q^2) \right]. 					\eqno (12)
$$
Taking into account that expression $\mu=\sqrt {Q}$ takes place in model
without self-gravitation let us write down growth rate (11) as:
$$
\gamma\simeq{\Omega_{in}\over 2} M {1
	+ Q^{-1} \over 1+Q^{-1/2}}(1-q^2). 			\eqno(13)
$$
Remarkable peculiarity of obtained result is extremely weak dependence
of growth rate on magnitude of density jump at $\rho_{in}\gg\rho_{ex}$, and
on the contrary proportionality to $Q^{-1/2} $ at $\rho_{in}\ll\rho_{ex}$.
Applying similar reasons we determine that for wave pattern angular
rotation phase speed $\Omega_p=Re\,\omega/m$ the following estimations
$\Omega_p\simeq\Omega_{in}$ at $\rho_{in}\gg\rho_{ex}$ and
$\Omega_p\simeq\Omega_{ex}$ at $\rho_{in}\ll\rho_{ex}$ are valid.
Hence the wave pattern rotation speed comes nearer to rotation speed
of more dense medium at significant density differences.

Note that in absence of speed shift ($q\equiv 1$) neutral solution
$\omega\simeq m\Omega_{in}$ respecting to non-growing disturbances
``frozen'' in gas follows from (11). Thus density kink alone cannot
cause instability.

To shed some light on physics of instability it is convenient to pass
from (11) to the dimensional form:
$$ \omega \simeq
{ M (\rho_{in}{c_s}_{in}\Omega_{in} + \rho_{ex}{c_s}_{ex}\Omega_{ex})
+ {i\over 2} (\rho_{in} + \rho_{ex}) R
(\Omega^2_{in} -\Omega^2_{ex} ) \over {c_s}_{in}\rho_{in}
+ {c_s}_{ex}\rho_{ex}}. 					\eqno (14)
$$
From (14) follows that the growth rate of considered instability
in the main order is equal to the ratio of difference of density
of centrifugal force, acting on a liquid particle fluctuating in
small vicinity of jump, to the sum of wave resistance
(characteristic impedances) of media on both sides from jump.
Thus this instability must be called centrifugal one,
just same as in case of disk of uniform density, which Morozov (1977)
had consider\footnote{\fta}{\ff Term ``centrifugal'' with reference
to considered instability was introduced later (Morozov et al.
1984, 1985a, 1985b).}.

\medskip
{\it b) An incompressible limit ($M \ll 1$).}

Taking into account $ | k_{in; ex} R | \propto M$ we make following
assumption: $|k_{in}R| \ll 1$, $|k_{ex}R| \ll 1$. Applied asymptotic
representation of modified Bessel functions at small values of argument
(Handbook of Mathematical Functions... 1964), we find the solution
(10) in the dimensionless and dimensional forms:

$$\eqalignno{
x \simeq & {1\over 1 + Q} \biggl\{m (q + Q) + (q-Q) \ \pm \cr
\pm& {\left[(q-Q)^2 - { 1\over 2} m (1-Q)^2 (1-q^2) -
	m^2Q (1-q^2) \right]}^{1/2} \biggl\}, 			& (15) \cr
} $$
$$\eqalignno{
\omega \simeq & {1\over\rho_{in} + \rho_{ex}}
	\biggl\{m (\Omega_{in}\rho_{in} + \Omega_{ex}\rho_{ex}) +
	(\Omega_{ex}\rho_{ex}-\Omega_{in}\rho_{in}) \ \pm \cr
\pm & \Bigl((\Omega_{ex}\rho_{ex} -\Omega_{in}\rho_{in})^2 -
	{1\over 2} m (\rho_{ex}-\rho_{in})^2 (\Omega^2_{in} -
	\Omega^2_{ex} ) \ -\cr
-&m^2\rho_{in}\rho_{ex} (\Omega_{in}-\Omega_{ex})^2 \Bigl)^{1/2} \biggl\}.
								&(16) \cr
} $$
The negativity of expression under square root sign in (15) and (16)
means instability.

In a case $Q = 1$ (uniform on density at $z = const$ disk) the result
received by Morozov (1977) follows from (15) :
$$ \omega \simeq {\Omega_{in}\over 2} \left\{m (q + 1) + (q-1) +
	i |1-q| \sqrt {m^2-1} \right\}. 			\eqno (17)
$$
In absence of velocity difference on jump ($q \equiv 1$) neutral
solutions follow from (15) as well as in the case of large
compressibility, but in this case they are the pair:
$\omega \simeq m\Omega_{in}$ and
$$ \omega \simeq \Omega_{in}\left(m
	+ 2 {1-Q\over 1 + Q} \right). 				\eqno (18)
$$
At significant density contrast (18) governs gyroscopic modes:
$\omega\simeq (m+2)\Omega_{in}$ at $Q\ll 1$ and
$\omega\simeq (m-2)\Omega_{in}$ at $Q\gg 1$. As one can see at $M\ll 1$
the density jump alone cannot cause instability too.

Let us analyse a physical nature of terms in (15) and (16) which
stabilize and destabilize the system.

For the disturbances with shortest azimuthal wavelength ($m\gg 1$)
$$ \omega \simeq m {\Omega_{ex}\rho_{ex}
	+ \Omega_{in}\rho_{in}\over \rho_{in} + \rho_{ex}} +
  im\,\sqrt {\rho_{in}\rho_{ex} {(\Omega_{in}-\Omega_{ex})^2\over (\rho_{in}
	+ \rho_{ex})^2}}. 					\eqno (19)
$$
follows from (16).

Introducing azimuthal wave number on jump $k_\varphi = m / R$ and
linear speeds of rotation inside and outside in immediate jump
proximity $V_{in} =R\Omega_{in}$ and $V_{ex} =R\Omega_{ex}$ accordingly,
it is easy to bring (19) to the form:
$$ \omega \simeq k_\varphi {\rho_{ex} V_{ex}
	+ \rho_{in} V_{in} \over\rho_{in} + \rho_{ex}} +
ik_\varphi {\sqrt {\rho_{in}\rho_{ex}} \over\rho_{in} +
	\rho_{ex}} {| V_{in} -V_{ex} |}.
								\eqno (20)
$$
The last expression is identical to classical expression
describing Kelvin--Helmholtz instability (KHI) developing
on vortex sheet in plainparallel current of an incompressible
fluid in absence of gravitation force ($g\equiv 0$):
$$ \omega = k_\parallel {\rho_1V_1 + \rho_2V_2\over\rho_1 + \rho_2}
\pm \sqrt {k_\parallel g {\rho_2-\rho_1\over\rho_1 + \rho_2} -
k^2_\parallel {\rho_1\rho_2 (V_1-V_2)^2\over (\rho_1 + \rho_2)^2}}.
                                                        	    \eqno (21)
$$
Here the index $1$ marks magnitudes concerning to the top layer,
$2$ to the bottom one respectively, $k_\parallel$ is wave number
of disturbances along vortex sheet\footnote{\fta}{\ff Inasmuch as
our model is quasi-two-dimensional ($\tilde v_z \equiv 0$),
equation (21) is written out in the assumption of concurrence
of directions {\bf V}$_1$, {\bf V}$_2$ and {\bf k}$_\parallel$.}.
Thus the nature of the last term under square root sign in (16)
always destabilizing jump surface is perfectly transparent:
it is caused by Bernoulli effect. This result is quite expected
as for disturbances with short on $\varphi$ wavelength effects
of curvature and rotation are the least essential.

Second term under square root sign in (16) can be rewritten as follows:
$$
  -k_\varphi {(\rho_{ex}-\rho_{in})^2\over (\rho_{ex}
	+ \rho_{in})^2} {R\Omega^2_{in} -R\Omega^2_{ex} \over 2}.
        		                                        \eqno (22)
$$
This term is directly proportional to centrifugal acceleration
difference on jump, renders stabilizing influence at
$\Omega_{ex} > \Omega_{in}$ and destabilizing at
$\Omega_{ex} <\Omega_{in}$. So the effects of centrifugal
stabilization and destabilization (see, for example, Nezlin \& Snezhkin
1990) take place, which are well known for a compressed fluid. However,
for an incompressible liquid they are shown for the first time.
It is also interesting that these effects act only in a case
$\rho_{in}\ne\rho_{ex}$. It has resulted to such fact that they were not
found by Morozov (1977). The latter, on our sight, is a direct
consequence of incompessibility and is explained as follows.

The incompressible fluid is characterized by significant (infinite as
an ideal) elasticity, because of that centrifugal effects cannot result
instability in the same way, as in compressible fluid, where liquid
particle releasing from vortex sheet by non-compensated centrifugal
or gravitational force is displaced on radius ``tightening'' particles
being ahead it. In given case the radial rearrangements of liquid
particles can come true only by their mutual replacement against
each other, as it occurs at development of Rayleigh--Taylor
instability, when particle of heavy fluid ``sinks'' on a place of
an emerging particle of light fluid. Not casually therefore similarity
of a considered combination (22) with first term under square root
sign in (21), just governing the development of Rayleigh--Taylor
instability on liquids interface in plainparallel flow. This similarity
becomes especially transparent in the case of significant density
contrast: assuming in (21) $\rho_1\gg\rho_2$ we get
$k_\parallel g$ and in the case $\rho_1\ll\rho_2$ this term gives
$-k_\parallel g$. From (22) as at $\rho_{in}\gg\rho_{ex}$ and at
$\rho_{in}\ll\rho_{ex}$ follows
$$
      -k_\varphi {R\Omega^2_{in} -R\Omega^2_{ex} \over 2}. 	\eqno (23)
$$
Thus in both cases we have longitudinal wave number and mass force
acceleration on vortex sheet production. Told leads to unexpected
conclusion that for considered axisymmetic vortex sheet a combination
of parameters
$$
   {\rho_{in}-\rho_{ex}\over\rho_{in} + \rho_{ex}} {R\Omega^2_{in}
	-R\Omega^2_{ex} \over 2}				\eqno (24)
$$
is equivalent $g$ in a plainparallel case.

Generalizing told we conclude that (22) describes buoyancy effects in
considered model and it is caused by buoyancy forces disbalance.

At last the first term under square root sign in (16) is easily
represented as
$$
   {| \rho_{ex} R^2\Omega_{ex}-\rho_{in} R^2\Omega_{in} | \over
	\rho_{ex} R^2 + \rho_{in} R^2},				\eqno (25)
$$
which is possible to interpete as the ratio of a half-difference of
density of an angular momentum on both jump sides to specific
(per volume unit) momentum of inertia of gas on jump. Tracing
genesis of this term in any case rendering stabilizing influence
upon jump it is possible to note that it is caused by action of
Coriolis' forces and reflects the tendency  to preserve an angular
momentum of system.

Rather interesting result is appearence of stability area on
parameter $q$ in a vicinity of  $q = 1$ at $Q\ne 1$. This area
wider as stronger $Q$ differs from unit and less as the less
is mode number $m$. The borders of stability constructed on
the asymptotic formula (15) for $m = 2, 3, 4$ are shown on Fig.~2.1.

On Fig.~2.2--2.5 results of the numerical solution of the
dispersion equation (11) by iterative Newton--Rafson method
are drawn. From them follows that the approximated formula (15)
will be good coordinated with dispersion curves in an incompressible
limit. The same good consent of results with (11) takes place
in a inverse case of essentially supersonic speed difference
on jump, if both $M^2 (1-q^2) \gg m$ and $M^2 (1-q^2) /\mu\gg m$
simultaneously are valid.

On our sight Fig.2.5 is rather interesting, on which stabilization
of centrifugal mode with growth $q$ is shown. Though it is occurs
formally at $M\gg 1$ the point of stabilization and branching on
two neutral modes corresponds to subsonic speed difference on jump
$M (1-q) /\mu \simeq 0.3 \ll 1$, therefore the mechanism of
stabilization is similar to described above.

Thus comparative analysis of results of numerical and asymptotic
analytical researches of the equation (10) allows to conclude
that all conclusions made in this section are valid.

\medskip 
{ \it 3. Conclusions.}
\medskip  

Let us summarize the basic conclusions concerning to stability of
non-homentopic angular speed and density discontinuity profiles of
finite thickness ($\xi\ll\Lambda_\Omega,\,\Lambda_\rho$, where
$\Lambda_\Omega,\,\Lambda_\rho$ are characteristic radial scales of
angular speed and density jumps) in model of uniform on pressure at
$z = const$ gaseous disk.

1. At essentially supersonic speed difference on jump centrifugal
instability is raised, which growth rate is equal to the ratio of
difference of density of centrifugal force acting on liquid particles
in vicinity of jump, to the sum of media wave resistances at
both sides of jump. Thus the angular rotation speed of a spiral
pattern in case of significant density drop is close to more
dense medium rotation speed.

2. At small compressibility of both media the Kelvin--Helmholtz
instability develops. It is caused by Bernoulli effect and
at presence of density difference also by buoyancy effects,
for which the role of external mass force density is played
by the following combination of parameters:
$$
      g_{eff} = {\rho_{in}-\rho_{ex}\over\rho_{in}
	+ \rho_{ex}} {R\Omega^2_{in} -R\Omega^2_{ex} \over 2}.
$$
Depending on sign of $g_{eff}$, an additional centrifugal
destabilization or stabilization of KHI takes place. Besides,
at comparison with KHI on plainparallel vortex sheet there are
the effects caused by angular momentum preservation. This effects
exert in any case a stabilizing influence, the more essential,
the larger is disturbance scale (and the less number $m$)
and the more considerable density drop.

3. Density drop alone in absence of speed difference is not capable
to result instability of considered model both in supersonic and
subsonic case.

It is necessary to note that as the KHI does not develop for
disturbances with wavelength comparable to or exceeding characteristic
scale $\Lambda_\Omega$ (that is modes with $m\gg1$ will appear neutral
in view of finite jump thickness) and the large-scale disturbances
are stabilized by virtue of conclusion 2, for smooth subsonic speed
jumps there is the area of stability on parameters $q,\,Q$.
The latter is necessary to take into account at stability analysis of
real objects .

The stabilization of Kelvin--Helmholtz instability for long wavelength
(with small $m$) disturbances of subsonic non-homentopic rotation
speed and density jump in view of suppression of this instability
for short-wave disturbances because of inevitably present finite
radial ``smearing'' of jump means an opportunity of rather
long-lasting existence of such jump. Thus the rotation speed on
radius of jump can be even supersonic. Such jumps are observed in
central areas of some flat galaxies (Afanesiev et al. 1988a, 1988b,
1988c, 1991, 1992) and can play a role of an internal reflecting
surface for waves amplifying on a corotation resonance (Fridman et al.
1994). Our analysis allows to assert that in case of significant
density contrast on jump the life-time of such surface will be
sufficient for development of resonant instability of the specified
waveguide layer. Otherwise the jump will be smeared out because of
``faster'' surface mode development earlier than resonant disturbances
will grow to nonlinear amplitudes.

\medskip
{\it Acknowledgement.} One of us (VVM) is grateful to INTAS
for support of this work by grant project N 95-0988.

\bigskip  
\centerline {\ref References}
\medskip  

Afanasiev, V.L., Burenkov, A.N., Zasov, A.V., \& Silchenko, O.K.,
	1988a, Afz., 28, 243 \par
Afanasiev, V.L., Burenkov, A.N., Zasov, A.V., \& Silchenko, O.K.,
	1988b, Afz., 28, 1040 \par
Afanasiev, V.L., Burenkov, A.N., Zasov, A.V., \& Silchenko, O.K.,
	1988c, Afz., 29, 155 \par
Afanasiev, V.L., Burenkov, A.N., Zasov, A.V., \& Silchenko, O.K.,
	1991, SvA.J., 68, 1134 \par
Afanasiev, V.L., Burenkov, A.N., Zasov, A.V., \& Silchenko, O.K.,
	1992, SvA.J., 69, 19 \par
Churilov, S.N. \& Shukhman, I.G., 1981, Astron. Tsircular, 1157, 1
	(in Russian) \par
Fridman, A.M., 1978, Sov. Phys. Uspekhi, 21, 536 \par
Fridman, A.M., \& Polyachenko, V.L., 1984, Physics of Gravitating
	Systems (New York: Springer-Verlag) \par
Fridman, A.M., 1989, in ``Dynamics of Astrophysical Discs'',
	ed. J.~Sellwood (Cambridge: Cambridge Univ. Press), p.185 \par
Fridman, A.M., \& Horuzhii, O.V., 1993, Sov. Phys. Uspekhi, 163, 79 \par
Fridman, A.M., Khoruzhii, O.V., Lyakhovich, V.V., Ozernoy, L.,
	\& Blitz, L., 1994, ASP Conference Series, 66, 285 \par
Goad, J.W., 1976, Ap. J. Suppl. Ser., 32, 89 \par
Gorbatsky, V.G., \& Serbin, V.M., 1983, Afz., 19, 79 \par
Gorbatsky, V.G., \& Usovitch, K.I., 1986, Afz., 25, 125 (in Russian) \par
Gorkavy, N.N., \& Fridman, A.M., 1994, Physics of planetary rings
	(Moscow: Science) \par
Handbook of Mathematical Functions... 1964, ed. M.~Abramowitz and
	I.~Stegun (National Bureau of Standards) \par
Haud, U.A., 1979, SvA. Lett., 5, 124 \par
Landau, L.D., 1944, Sov. Phys. Dokl., 44, 151 \par
Morozov, A.G., Fainstein, V.G., \& Fridman, A.M., 1975,
	in ``Dynamics and evolution of star systems'',
	VAGO, Moscow--Leningrad, 238 p. (in Russian) \par
Morozov, A.G., Fainstein, V.G., \& Fridman, A.M., 1976a, JETP, 44, 653 \par
Morozov, A.G., Fainstein, V.G., \& Fridman, A.M., 1976b,
	Sov. Phys. Dokl., 21, 661 \par
Morozov, A.G., 1977, SvA. Lett., 3, 103 \par
Morozov, A.G., 1979, SvA. J., 23, 27 \par
Morozov, A.G., Nezlin, M.V., Snezhkin, E.N., \& Fridman, A.M., 1985a,
	Phys. Lett., 109A, 228 \par
Morozov, A.G., Nezlin, M.V., Snezhkin, E.N., \& Fridman, A.M., 1985b,
	Sov. Phys. Uspekhi, 28, 101 \par
Morozov, A.G., 1989, KFNT, 5, 75 (in Russian) \par
Morozov, A.G. \& Mustsevoy, V.V., 1989, Preprint of Volgograd State
	Univ., 5-89 \par
Nezlin, M.V., \& Snezhkin, E.N., 1993, Rossby vortices, spiral structures,
	solitons (Berlin: Springer--Verlag) \par
Rubin, V.C., \& Ford, W.K., 1970, Ap. J., 159, 379 \par
Rubin, V.C., Burstein, D., Ford, W.K., \& Thonnard, N., 1985,
	Ap. J., 81, 289 \par
Sanders, P.B., Solomon, P.M., \& Scoville, N.Z., 1984, Ap. J., 276, 182 \par
Sinha, R.R., 1978, A\&A., 69, 227 \par
Torgashin, Yu.M., 1986, PhD diss., Tartu (in Russian) \par
Usovitch, K.I., 1988, Afz., 28, 510 \par

\vfil
\eject

\bigskip
\epsfxsize=\hsize
\epsfbox{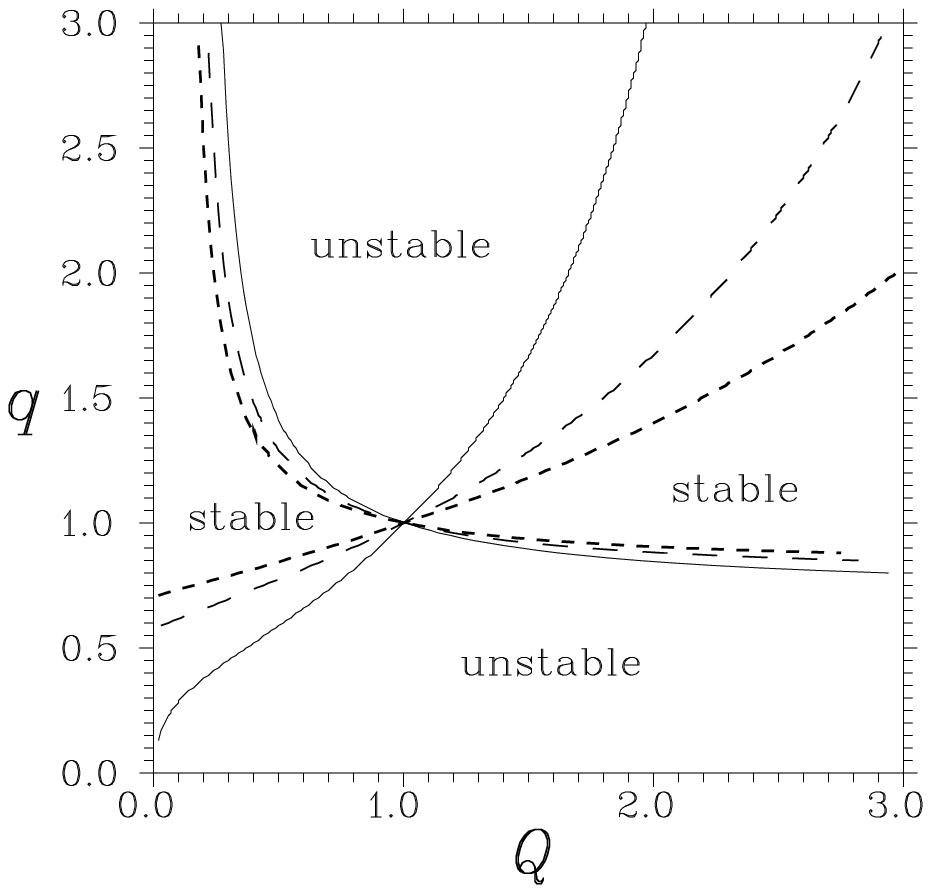}
\bigskip

{\bf Fig.2.1.} Border of marginal stability of azimuthal modes
$m=2,3,4$ (accordingly, continuous line, long dash, short dash)
according to the asymptotic formula (15).

\vfil
\eject

\bigskip
\epsfxsize=\hsize
\epsfbox{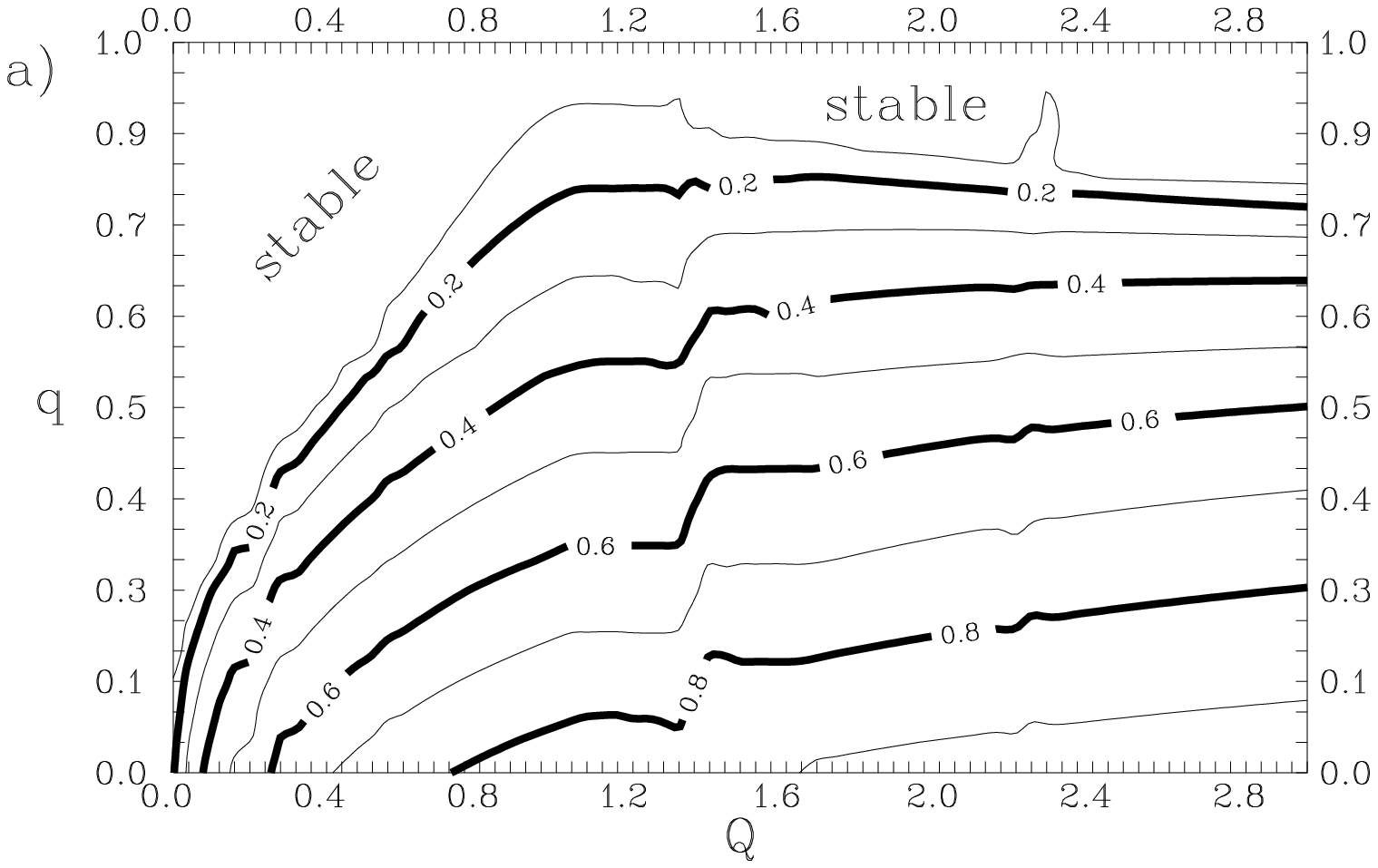}
\medskip
\epsfxsize=\hsize
\epsfbox{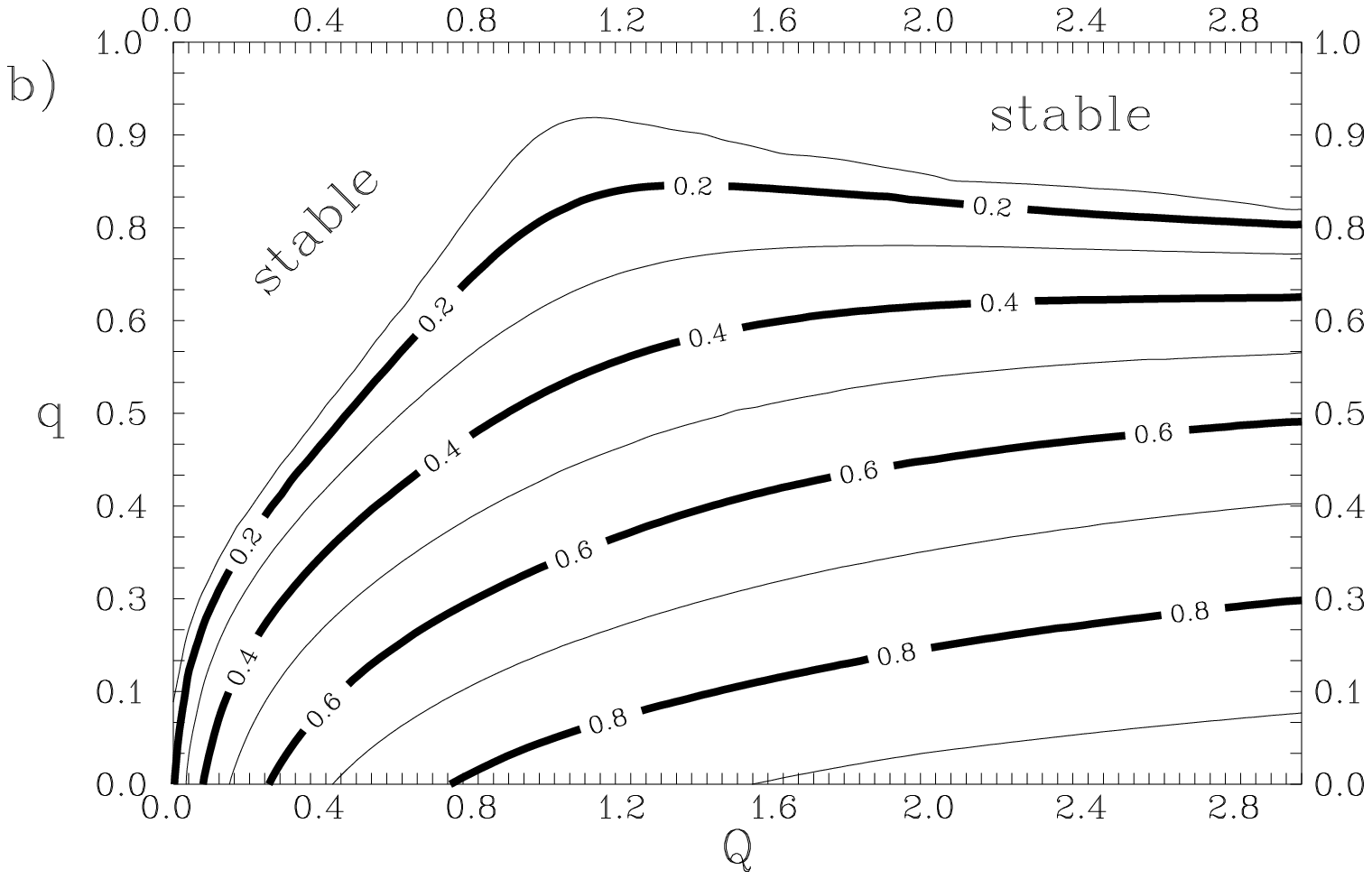}
\bigskip

{\bf Fig.2.2.} Lines of a level of dimensionless growth rate
for $m=2$: (a) on the data of the numerical solution (10),
(b) according to asymptotic (15). The breaks of isolines on the
top drawing are probably artefact of the numerical solution.

\vfil
\eject

\bigskip
\epsfxsize=\hsize
\epsfbox{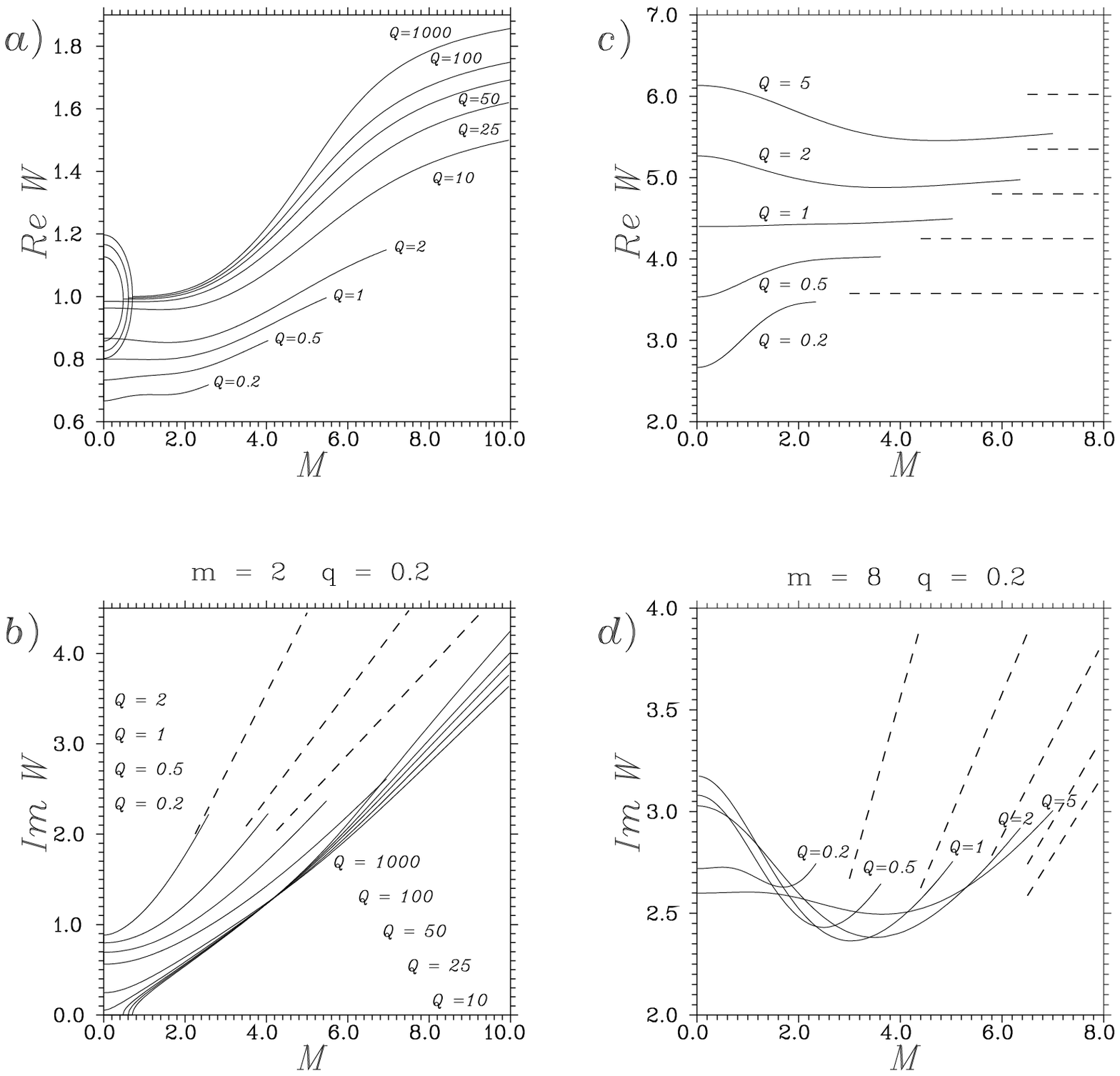}
\bigskip

{\bf Fig.2.3.} Dependence of dimensionless frequency
$Re\,\omega/\Omega_{in}$ (a), (c) and dimensionless growth rate
$Im\,\omega/\Omega_{in}$ (b), (d) from Mach number on the data of
the numerical solution (10) (continuous lines) and
asymptotics (11) (dashed lines) for various sets of
parameters $m,\ q,\ Q$.

\vfil
\eject

\bigskip
\epsfxsize=\hsize
\epsfbox{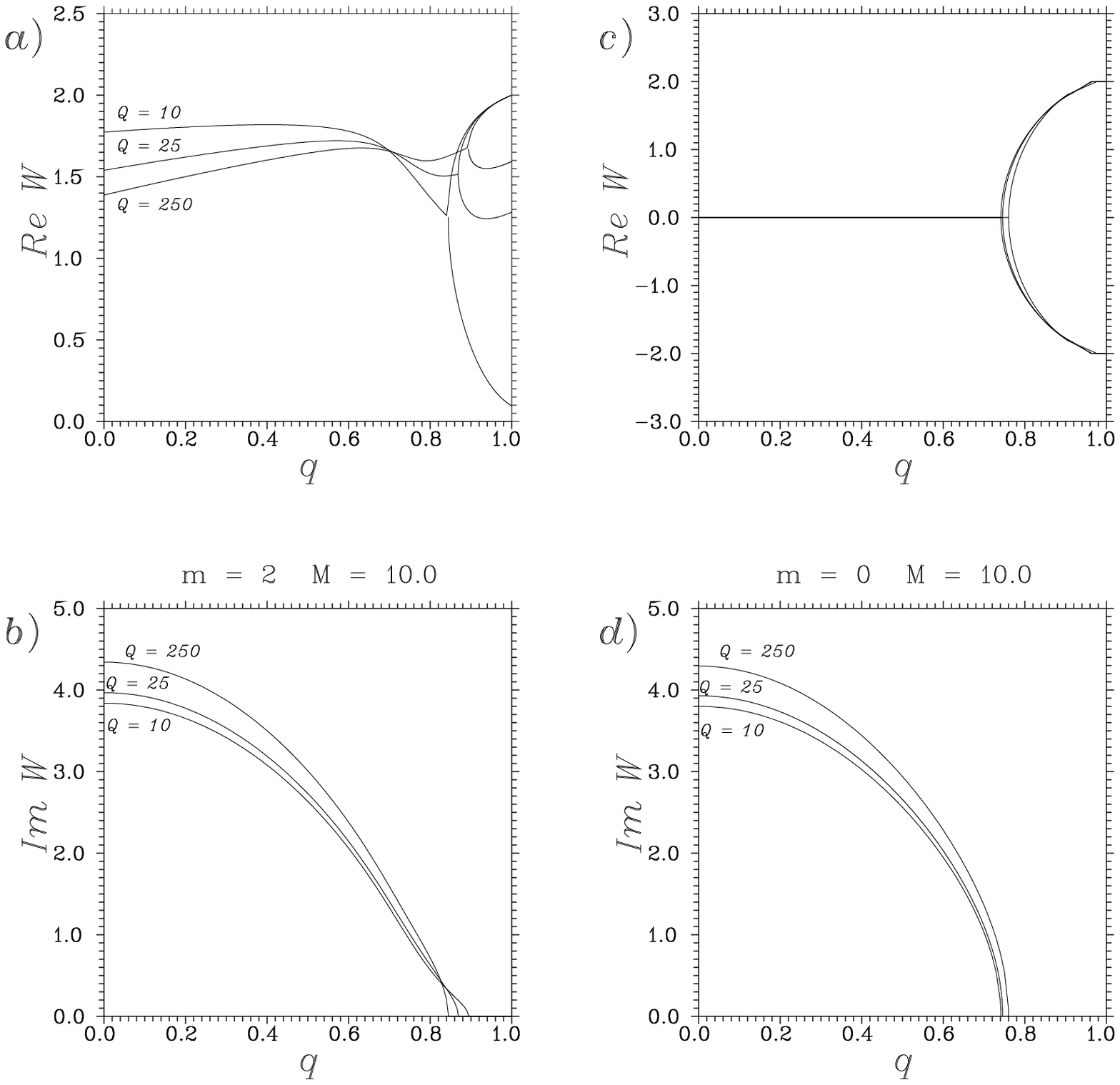}
\bigskip

{\bf Fig.2.4.} Dependence of dimensionless frequency (a), (c)
and dimensionless growth rate (b), (d) from relative density difference
on jump on the data of the numerical solution (10) (continuous lines)
and asymptotics (11) (dashed lines) for various sets of parameters:
$m,\ q,\ M$.

\vfil
\eject

\bigskip
\epsfxsize=\hsize
\epsfbox{2-5.eps}
\bigskip

{\bf Fig.2.5.} Dependence of dimensionless frequency (a), (c)
and dimensionless growth rate (b), (d) from value of speed difference
on jump on the data of the numerical solution (10) for
non-axisymmetric ($m=2$) and axisymmetic ($m=0$) modes.

\vfil
\eject

\end